\documentclass[useAMS,usenatbib,usegraphicx]{mn2e}

\title[ULIRGs and HLIRGs in the SDSS, 2dFGRS and 6dFGS]
{Ultraluminous and Hyperluminous Infrared Galaxies in the
SDSS, 2dFGRS and 6dFGS}
\author[H. S. Hwang et al.]{Ho Seong Hwang,$^{1,2}$\thanks{E-mail:
hshwang@astro.snu.ac.kr} Stephen Serjeant,$^{3,4}$ Myung Gyoon
Lee$^{1}$,
\newauthor Kang Hwan Lee,$^{3,5}$ and Glenn White$^{4,6}$\\
$^{1}$Astronomy Program, Department of Physics and Astronomy, FPRD, Seoul National University, Seoul
151-742, Korea \\
$^{2}$Visiting Research Associate,
Centre for Astrophysics \& Planetary Science, School of
Physical Sciences, University of Kent, \\
Canterbury, Kent CT2 7NR \\
$^{3}$Centre for Astrophysics \& Planetary Science, School of
Physical Sciences, University of Kent, Canterbury, Kent CT2 7NR \\
$^{4}$Astrophysics Group, Department of Physics and Astronomy, The
Open University, Milton Keynes MK7 6AA \\
$^{5}$National Science Museum Planning Office, Ministry of Science
\& Technology, Gwacheon Kyungi, 427-715, Korea \\
$^{6}$CCLRC Rutherford Appleton Laboratory, Chilton, Didcot,
Oxfordshire OX11 0QX}

\begin{document}


\pagerange{\pageref{firstpage}--\pageref{lastpage}} \pubyear{2002}

\maketitle

\label{firstpage}

\begin{abstract}

We present a result of cross-correlating the {\it Infrared
Astronomical Satellite} Faint Source Catalogue ({\it IRAS\/} FSC)
with the spectroscopic catalogues of galaxies in the Fourth Data
Release of Sloan Digital Sky Survey (SDSS), the Final Data Release
of 2dF Galaxy Redshift Survey (2dFGRS) and the Second Data Release
of 6dF Galaxy Survey (6dFGS). We have identified 324 ultraluminous
infrared galaxies (ULIRGs) including 190 newly discovered ULIRGs,
and 2 hyperluminous infrared galaxies (HLIRGs). Adding these new
ULIRGs, we increase the number of known ULIRGs by about 30 per
cent. The reliability of the cross-correlation is estimated using
the likelihood ratio method. The incompleteness of our sample
introduced by the identification procedure in this study is estimated
to be about 5 per cent. Our sample covers the redshift range of
z=0.037$-$0.517 with a median redshift of $\bar{z}$=0.223, which
is larger than that ($\bar{z}$=0.184) of the sample of previously
known ULIRGs.
\end{abstract}


\section{Introduction}

The interest in luminous infrared galaxies, in particular 
ultraluminous infrared galaxies\footnote{We define ULIRGs as 
the galaxies whose infrared (1--1000 $\umu$m) luminosities are
greater than 10$^{12}$ L$_\odot$.}
(ULIRGs) and the hyperluminous
infrared galaxies\footnote{We define HLIRGs as the galaxies whose 
infrared luminosities are greater than 10$^{13}$
L$_\odot$.}
(HLIRGs), has been growing since the launch of {\it
Infrared Astronomical Satellite} ({\it IRAS}; \citealt{neu84}) in 1983.
A great deal of effort has been made to understand the origin of
the enormous infrared luminosities of these populations and the time
evolution of individual ULIRGs, which are summarised in \citet{sm96} and
\citet{lonsdale06}.
It is generally accepted that 
dust heated by some combination of starburst and
active galactic nuclei (AGN) activity is responsible for the IR
luminosity (e.g. \citealt{far03}). 
However, it is not yet clear which is the dominant power source. 
There are several suggestions for the time evolution of individual ULIRGs.
\citet{san88a} suggested that ULIRGs may be an
initial, dust-shrouded stage of optical QSOs, while \citet{far01}
proposed that ULIRGs are not a simple transition stage from galaxy
mergers to QSOs but evolve through diverse paths.
On the other hand, several authors suggested that ULIRGs might
evolve into moderately massive (L$^*$) field ellipticals
\citep{ks92,genzel01,tac02}.

Infrared space telescopes have driven much recent progress
(see \citealt{genzel00} and \citealt{verma05} for a review). 
\citet{genzel98} showed that several diagnostic lines
in mid-infrared spectra using {\it Infrared Space Observatory}
({\it ISO}; \citealt{kessler96}) are very powerful in characterizing
the power source of ULIRGs. They concluded that most ULIRGs (80\%)
are powered by starbursts, but at least half of their sample
require both a starburst and an AGN. Similar results are found in
\citet{lutz98} and \citet{rig99}. Moreover, \citet{tran01} showed
that ULIRGs with L$_{ir}<$10$^{12.4}$L$_\odot$ are mostly
starburst dominated, while ULIRGs with
L$_{ir}>$10$^{12.4}$L$_\odot$ more likely to contain AGN. The {\it
Spitzer Space Telescope} \citep{werner04} and submillimetre/millimetre-wave
cameras such as {\it SCUBA} are now extending
samples of ULIRGs to higher redshifts ($z>1$). Several authors
\citep{houck05,lutz05,yan05} took spectra of optically faint and
infrared luminous population using {\it Spitzer}, detecting broad
spectral features such as PAH (Polycyclic Aromatic Hydrocarbon)
emission and silicate absorption. They showed that the majority of
these population are at high redshift ($z\sim2$) and have
mid-infrared spectral shapes similar to local AGN-dominated
ULIRGs.

Since the publication of the {\it IRAS} Point Source Catalogue
(\citealt{psc88}; hereafter PSC) and the {\it IRAS} Faint Sources
Catalogue -- Version 2 (\citealt{mos92}, hereafter FSC92), many 
wide-area redshift survey follow-up campaigns have been conducted.
In total, these have led to a heterogeneous compilation of a few
hundred ULIRGs from sources such as the
IRAS 1.2 Jy Redshift Survey \citep{strauss90,strauss92,fisher95},
the IRAS 1 Jy Survey of ULIRGs \citep{ks98}, the QDOT all-sky IRAS
galaxy redshift survey \citep{law99}, the Point Source Catalogue
redshift survey \citep{saun00}, the FIRST/FSC sample
\citep{stanford00}, and the Revised Bright Galaxy
Sample\footnote{The earlier version of this survey is the Bright
Galaxy Survey \citep{soifer86,soifer87,soifer89,sanders95}.}
\citep{san03}. The small area surveys are summarised in
\citet{sm96} and \citet{lonsdale06}. The majority of ULIRGs found
in all sky redshift surveys are from the nearby universe ($z<0.5$) and
have abundant multi-wavelength data. Such objects are useful prototypes
for the study of high redshift infrared luminous galaxies. 
However, these redshift surveys are still far from being a complete
spectroscopic survey of all {\it IRAS}
sources. Therefore, many ULIRGs remain undiscovered in FSC92.

More recent galaxy redshift surveys such as the Sloan Digital
Sky Survey (SDSS, \citealt{york00}), the 2dF Galaxy Redshift Survey
(2dFGRS, \citealt{col01}) and the 6dF Galaxy Survey (6dFGS,
\citealt{jones04}) have provided redshifts for much larger samples of
galaxies. These data will clearly be very useful for 
finding new ULIRGs in FSC92, which has been attempted by several
studies 
\citep{cao06,goto05,pasq05}. However, \citet{goto05} and
\citet{pasq05} cross-correlated the {\it IRAS\/} sources with the
SDSS spectroscopic sample of galaxies using a circular matching
tolerance instead of using a positional error ellipse of the {\it IRAS\/} sources. 
\citet{cao06} performed a similar cross-correlation using the
formal positional error ellipses of the {\it IRAS\/} sources,
but they used only the Second Data Release of SDSS.

In this paper we present a search for ULIRGs and
HLIRGs from a cross-correlation of FSC92 with the most recent
spectroscopic catalogues of galaxies in SDSS, 2dFGRS and 6dFGS,
using positional matches in the {\it IRAS\/}  error ellipses.
The data used for the cross-correlation and the identification
algorithm for ULIRGs and HLIRGs are presented in Section 2. Our
main results are presented in Section 3, and are discussed in Section
4. We summarize and conclude in Section 5. Throughout this
paper, we assume that ${H_0}$ = 70 km s$^{-1}$ Mpc$^{-1} $,
$\mathrm {\Omega_0} $ = 0.3 and $\mathrm{\Lambda}$ = 0.7.

\section[]{Identification of ULIRGs and HLIRGs}
\subsection {The Samples and Cross-Correlation}\label{sample}
We cross-correlated the {\it IRAS\/} sources in FSC92 with the
spectroscopic sample of galaxies in the Fourth Data Release of SDSS
(\citealt{am05}, hereafter SDSSDR4), the Final Data Release of
2dFGRS (\citealt{col01}, hereafter 2dFGRSDRF) and the Second Data
Release of 6dFGS (\citealt{jones05}, hereafter 6dFGSDR2). The
FSC92 contains about 173,000 infrared sources with measured fluxes
at 12 $\umu$m, 25 $\umu$m, 60 $\umu$m and 100 $\umu$m. The SDSSDR4
and the 2dFGRSDRF have a similar median depth of $\bar{z}$=0.11.
The SDSSDR4 contains about 473,000 galaxies over 4,800 deg$^2$,
and the 2dFGRSDRF contains about 246,000 galaxies over 2,000
deg$^2$. The 6dFGSDR2 which contains about 89,000 galaxies, covers
a wider area ($\sim$13,600 deg$^2$) than the other surveys, but
has a shallower median depth of $\bar{z}$=0.05.

The positional uncertainties of the {\it IRAS\/} sources are
different for the in-scan direction and the cross-scan direction
(typically 5$''$ for the in-scan direction and 16$''$ for the
cross-scan direction, 1 $\sigma$), and vary from source to source 
(1$''-$13$''$ for the in-scan direction, or minor axis of
positional uncertainty ellipse and 3$''-$55$''$ for the cross-scan
direction, or major axis of positional uncertainty ellipse, 1
$\sigma$). The positional
uncertainties of the optical identifications themselves are negligible
in comparison. We use the positional uncertainty ellipse of FSC92 to
find matching counterparts of the {\it IRAS\/} sources, instead of
using a circular matching tolerance (to be discussed in Section
\ref{rel}). If a galaxy in a redshift survey lies within 3
$\sigma$ positional uncertainty ellipse of the {\it IRAS\/}
source, we regard it a match. As a result of this matching, we 
find that 8382, 2091
and 10197 {\it IRAS\/} sources have optical
counterparts in the SDSSDR4, 2dFGRSDRF and 6dFGSDR2, respectively.
Some {\it IRAS} sources (615 for SDSSDR4, 201 for 2dFGRSDRF, and
91 for 6dFGSDR2) have more than two galaxies within 3 $\sigma$
error ellipse.
In these cases, we computed the ``Likelihood Ratio (LR)''
(\citealt{sutherland92}, to be discussed in \ref{rel}) of each
association, and selected a more appropriate galaxy with the larger value of
LR as an optical counterpart. In total, we compiled a
list of 19,380 sources by collecting all matched sources from
our disparate galaxy redshift surveys.

\begin{table*}
\caption{ULIRGs found in SDSSDR4, 2dFGRSDRF and
6dFGSDR2\label{tab-ULIRG1}}
\begin{tabular}{clllllc}
\hline\hline FSC & SDSS & & 6dFGS & &
2dFGRS & \\
\cline{2-2} \cline{4-4} \cline{6-6} NAME &
ID~~~~~~~~~~~~~~~~~~~~~~~~~~~z~~~~~~~zconf & &
ID~~~~~~~~~~~~~~~~~~~~~~z~~~~~~~Q &  &
ID~~~~~~~~~~~~~~~~z~~~~~~~Q & Final \\
\hline
F00050$-$3259* &                     ~      ~       & &                 ~          & &   TGS434Z061~ 0.285~ 4 &  3 \\
F00091$-$0738~ &                     ~      ~       & &  g0011433$-$072207~ 0.118~ 4 & &             ~          &  2 \\
F00091$-$3905* &                     ~      ~       & &  g0011432$-$384913~ 0.254~ 3 & &             ~          &  2 \\
F00095$-$5948* &                     ~      ~       & &  g0011586$-$593133~ 0.235~ 4 & &             ~          &  2 \\
F00184$-$3331* &                     ~      ~       & &  g0020557$-$331428~ 0.239~ 3 & &   TGS497Z156~ 0.239~ 4 &  2 \\
F00256$-$0208~ &                     ~      ~       & &  g0028143$-$015146~ 0.277~ 3 & &             ~          &  2 \\
F00285$-$3140~ &                     ~      ~       & &  g0030593$-$312445~ 0.218~ 2 & &   TGS439Z175~ 0.217~ 4 &  3 \\
F00318$-$3137* &                     ~      ~       & &  g0034133$-$312118~ 0.285~ 4 & &             ~          &  2 \\
F00335$-$2732~ &                     ~      ~       & &                 ~          & &   TGS206Z015~ 0.069~ 5 &  3 \\
F00406$-$3127~ &                     ~      ~       & &  g0043032$-$311050~ 0.343~ 1 & &             ~          &  2 \\
\hline
\end{tabular}
\begin{flushleft}
Column descriptions: {\it (1) } The {\it IRAS\/} object name in
the FSC92. Asterisks represent the sources that are identified as
ULIRGs for the first time in this study; {\it (2-4) } The
identification, the redshift, and the redshift confidence value in
SDSSDR4; {\it (5-7) } The identification, the redshift, and the
redshift quality parameter in 6dFGSDR2; {\it (8-10) } The
identification, the redshift,
and the redshift quality parameter in 2dFGRSDRF; {\it (11) } Finally
accepted optical counterpart (1$-$SDSSDR4, 2$-$6dFGSDR2, 3$-$2dFGRSDRF).
\end{flushleft}
\end{table*}

\begin{table*}
\caption{The {\it IRAS} data for the Final Sample of ULIRGs\label{tab-ULIRG2}}
\begin{tabular}{crrrrrrc}
\hline\hline FSC & RA & Dec & 12$\umu$m & 25$\umu$m & 60$\umu$m &
100$\umu$m & Flux \\
NAME & (J2000) & (J2000) & (Jy) & (Jy) & (Jy) & (Jy)  & Qual\\
\hline
F00050$-$3259* & 00 07 34.6 & $-$32 43 03 &   0.066 (0.018)&   0.144 (0.042)&   0.222 (0.045)&   0.758 (0.220)& 1131 \\
F00091$-$0738~ & 00 11 43.3 & $-$07 22 05 &   0.071 (0.020)&   0.215 (0.054)&   2.626 (0.184)&   2.521 (0.202)& 1232 \\
F00091$-$3905* & 00 11 42.3 & $-$38 49 15 &   0.106 (0.030)&   0.125 (0.038)&   0.316 (0.047)&   0.756 (0.227)& 1131 \\
F00095$-$5948* & 00 11 58.8 & $-$59 31 28 &   0.074 (0.014)&   0.052 (0.010)&   0.313 (0.034)&   0.681 (0.123)& 1132 \\
F00184$-$3331* & 00 20 57.7 & $-$33 14 28 &   0.105 (0.028)&   0.120 (0.035)&   0.334 (0.047)&   0.613 (0.141)& 1132 \\
F00256$-$0208~ & 00 28 14.4 & $-$01 51 47 &   0.108 (0.031)&   0.323 (0.090)&   0.602 (0.060)&   0.611 (0.147)& 1132 \\
F00285$-$3140~ & 00 31 03.0 & $-$31 24 18 &   0.144 (0.042)&   0.141 (0.041)&   0.389 (0.051)&   0.688 (0.199)& 1132 \\
F00318$-$3137* & 00 34 16.0 & $-$31 21 04 &   0.108 (0.030)&   0.167 (0.048)&   0.257 (0.062)&   0.563 (0.158)& 1132 \\
F00335$-$2732~ & 00 35 59.2 & $-$27 15 42 &   0.144 (0.045)&   0.632 (0.069)&   4.294 (0.472)&   3.207 (0.225)& 1332 \\
F00406$-$3127~ & 00 43 03.0 & $-$31 10 53 &   0.060 (0.016)&   0.091 (0.025)&   0.717 (0.057)&   0.994 (0.169)& 1132 \\
\hline
\end{tabular}
\begin{flushleft}
Column descriptions: {\it (1) } The {\it IRAS\/} object name in
the FSC92. Asterisks represent the sources that are identified as
ULIRGs for the first time in this study; {\it (2-3) } Right
ascension and declination; {\it (4-7) } The {\it IRAS\/} flux
density (and its uncertainty)
at 12$\umu$m, 25$\umu$m, 60$\umu$m and 100$\umu$m; {\it
(8)} The {\it IRAS\/} flux density qualities at each band (1$-$upper
limit, 2$-$moderate quality, 3$-$high quality).
\end{flushleft}
\end{table*}

\begin{table}
\begin{center}
\caption{The Final Sample of ULIRGs\label{tab-ULIRG3}}
\begin{tabular}{cccrl}
\hline\hline FSC & Final & log$L_{ir}$ & log(LR) &Class  \\
NAME & z & ($L_\odot$) & & L/S \\
\hline
F00050$-$3259* & 0.2855 & 12.12 &   0.5042 &  7/V    \\
F00091$-$0738~ & 0.1178 & 12.30 &   0.6345 &  7/IV   \\
F00091$-$3905* & 0.2535 & 12.14 &   0.5944 &  7/V    \\
F00095$-$5948* & 0.2349 & 12.06 &   0.6102 &  7/IV   \\
F00184$-$3331* & 0.2387 & 12.10 &  $-$0.1065 &  7/V    \\
F00256$-$0208~ & 0.2770 & 12.52 &   0.5735 &  5/IIIa \\
F00285$-$3140~ & 0.2166 & 12.07 &  $-$0.9137 &  4/I    \\
F00318$-$3137* & 0.2846 & 12.18 &   0.3252 &  7/V    \\
F00335$-$2732~ & 0.0686 & 12.01 &   0.3877 &  7/IV   \\
F00406$-$3127~ & 0.3425 & 12.82 &   0.5803 &  7/V    \\
\hline
\end{tabular}
\end{center}
\begin{flushleft}
Column descriptions: {\it (1) } The {\it IRAS\/} object name in
the FSC92. Asterisks represent the sources that are identified as
ULIRGs for the first time in this study;
  {\it (2) } Finally accepted redshift from galaxy redshift survey;
  {\it (3) } Infrared luminosity calculated in this study;
  {\it (4) } Likelihood ratio;
  {\it (5) } Morphological class using Lawrence classification (L) and
Surace classification (S). See the description in the end of this section.
\end{flushleft}
\end{table}

\subsection{Selection Criteria}\label{identi}
We used the following steps in order to identify bona-fide ULIRGs
and HLIRGs from the list of all matched sources:
\begin{enumerate}
\item For 19,380 sources in the list of all matched sources, we
calculated the IR luminosities using the {\it IRAS\/} 60 $\umu$m
fluxes and the redshifts of galaxies on the
assumption that the ULIRGs and HLIRGs have M82 starburst Spectral
Energy Distributions (SEDs). A substantial fraction of the
{\it IRAS} sources have upper limit fluxes at 12 $\umu$m, 25
$\umu$m and 100 $\umu$m, so we use this single-band selection to
obtain a less heterogeneous final catalogue
(to be discussed in Section \ref{rel}).
\label{crit1}

\item We selected the sources for which their 60 $\umu$m flux qualities are
``high'' or ``moderate'' \citep{mos92} in order to restrict our
analysis to those 
sources with reliable 60 $\umu$m detections\footnote{The
flux measurements of high quality, moderate quality or upper
limits are represented as number of 3, 2, 1 respectively in
FSC92.}, obtaining 19,335 sources.\label{crit2}

\item Of the 19,335 reliable sources, we classified 
483 sources as ULIRG candidates (IR
luminosities greater than 10$^{12}$ L$_\odot$), and classified
14 sources as HLIRG candidates 
(IR luminosities are
greater than 10$^{13}$ L$_\odot$).\label{crit3}

\item We then assessed the quality of the redshift for the ULIRG 
and HLIRG candidates. 
We required a redshift confidence of $\ge$ 0.65
for the SDSSDR4, a redshift quality of $\ge$ 3 for the
2dFGRSDRF, and a redshift quality is equal to 3 or 4 for the
6dFGSDR2. See each data release paper for the descriptions of
these redshift quality parameters. On this basis, 
we secured 333 ULIRG and no HLIRG candidates with acceptable
redshift qualities.\label{crit4}

\item We used the NASA/IPAC Extragalactic Database (NED) Near
Position Search in order to reject chance coincidences of {\it
IRAS} sources for which optical counterparts are already listed in
the literature. We rejected 13 ULIRG candidates (two sources moved
to a list of HLIRGs), and added 4 sources (to be discussed in
Section \ref{rel}). Our redshifts for these four are confirmed in the
literature, although their redshift qualities do not satisfy our
criteria. Finally we found 324 ULIRGs and 2 HLIRGs (the final two
HLIRGs will be discussed in Section \ref{HLIRGs}).\label{crit5}

\end{enumerate}
Table \ref{tab-ULIRG1} lists the final sample of 324 ULIRGs with
the names of the {\it IRAS\/} sources, the identifications of
their optical counterparts, redshifts and redshift quality
parameters in each galaxy redshift survey. In some cases, more
than two disparate galaxies from different galaxy redshift surveys
are located within an error ellipse of one {\it IRAS} source.
Therefore, we present the flags indicating which galaxy has been 
adopted as the optical counterpart in the final column of Table
\ref{tab-ULIRG1}. We present the {\it IRAS} FSC92 data in Table
\ref{tab-ULIRG2}. Adopted redshifts, IR luminosities, log
(LR) and morphological classes (to be discussed at the end of this
section) are listed in Table \ref{tab-ULIRG3}.

\begin{figure*}
\begin{center}
\includegraphics [width=165mm] {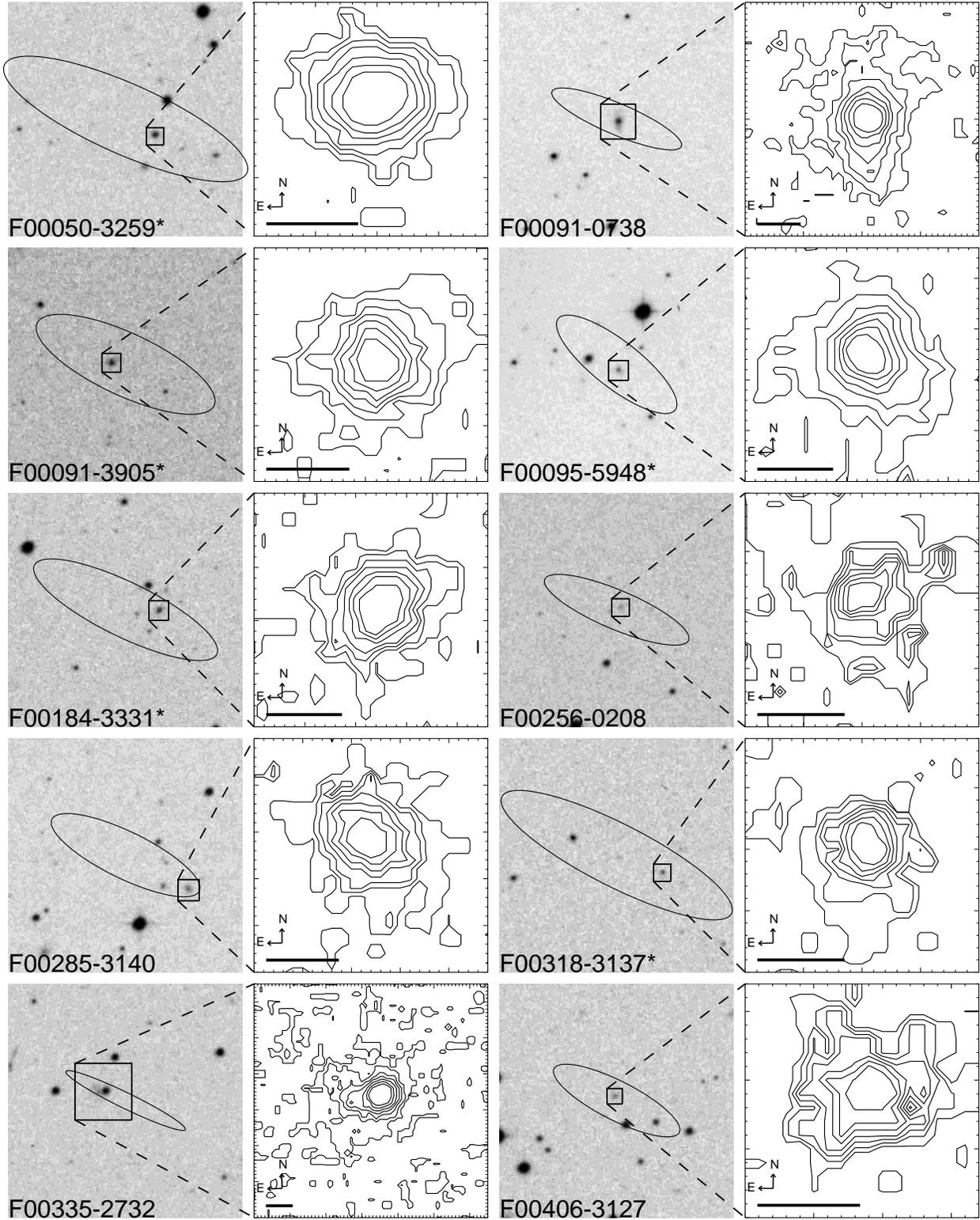}
\end{center}
\caption{The 3$'\times$3$'$ finding charts (left in each panel)
and contour plots (right in each panel) of 324 ULIRGs identified
in this study. The finding charts centred on {\it IRAS} source
positions are extracted from the SDSS $gri$ composite images or
the SuperCOSMOS scans of r$_F$ survey plates. North is up, and
the East is to the left. The ellipse represents 3$\sigma$
{\it IRAS} 
positional uncertainty. The optical counterpart in the redshift
surveys is marked by square indicating the size, orientation and
location of the contour plot in the finding chart. The {\it
IRAS\/} object names are presented in the bottom left corner, and
asterisks represent the newly identified ULIRGs in this study. The
contour plots centred on optical counterparts represent the
intensities of ULIRGs in the SDSS r-band images or the SuperCOSMOS
r$_F$ images. The size of each contour plot is 40 kpc $\times$ 40 kpc.
The contour levels increase from the sky level ($I_{\rm sky}$) to the peak value ($I_{\rm max}$)
by factors $10^{(I_{\rm max}-I_{\rm sky})/8}$.
The orientation is indicated by the 
arrows and the thick, horizontal bar represents 5 arcsec in each
contour plot.} \label{fig-ULIRGs}
\end{figure*}

Figure \ref{fig-ULIRGs} shows a
sample of 3$'\times$3$'$ finding charts (left hand panels) for
the final sample of ULIRGs extracted from the SDSS $gri$ composite
images or the SuperCOSMOS scans of r$_F$ survey plates.
The contour plot for each ULIRG using the SDSS r-band images or
the SuperCOSMOS r$_F$ images is also presented in Figure
\ref{fig-ULIRGs} (right hand panels). Using Figure
\ref{fig-ULIRGs}, we classify the interaction types of the ULIRGs
in the final sample. The ULIRGs are classified according to the
modified version of the interaction classification schemes of
\citet[hereafter Lawrence classification]{law89} (see, e.g.,
\citealt{far01}) and of \citet[hereafter Surace
classification]{sur98} (see, e.g., \citealt{vks02}), as are
summarised in Table \ref{tab-class}. The results of the
classification are given in the last column of Table
\ref{tab-ULIRG3}. The complete versions of Table \ref{tab-ULIRG1},
Table \ref{tab-ULIRG2}, Table \ref{tab-ULIRG3} and Figure
\ref{fig-ULIRGs} are available only in the electronic issue. A
brief discussion of some ULIRGs is given in Appendix A 
that is available in the electronic issue.

\begin{table*}
\centering
\caption{The interaction classification scheme of ULIRGs used in
this study \label{tab-class}}
\begin{tabular}{cll}
\hline\hline Scheme & Class & Description \\ \hline
Lawrence & 0 & Isolated source with no signs of interaction or merging \\
classification & 1 &  Source with a faint companion (2$\sim$4 mag
fainter than the
source) in the range of 40$\sim$200 kpc \\
 & 2 & Source with a bright companion (less than 2 mag fainter than
the source) in the range of 40$\sim$200 kpc \\
 & 3 & Source with a faint companion less than 40 kpc \\
 & 4 & Source with a bright companion less than 40 kpc \\
 & 5 & Source interacting with a companion and showing signs of
interaction \\
 & 6 & Merger/More than two nuclei in common envelope \\
 & 7 & Merger/Single nuclei in common envelope \\
\hline

Surace & I & First approach -- Unperturbed and separated galaxies
with no signs of interaction or merging\\
classification & II & First contact -- Overlapped disks but no
strong bars and
tidal tails\\
 & IIIa & Pre-merger a -- Two recognisable nuclei with strong signs
of interaction and separated more than 10 kpc\\
 & IIIb & Pre-merger b -- Two recognisable nuclei with strong signs
of interaction and separated less than 10 kpc\\
 & IV & Merger -- Only one nucleus seen with strong tidal features\\
 & V & Old merger -- No direct signs of tidal features but disturbed
central morphology \\
\hline
\end{tabular}
\end{table*}

\subsection{Reliability of Identifications}\label{rel}
A variety of definitions of ULIRGs have been used in the literature,
and depend on the minimum luminosity, the adopted spectral energy
distributions and the cosmology ($L_{8-1000} \ge 10^{12} $L$_\odot$
with $H_0$ =75 km s$^{-1}$Mpc$^{-1}$ and $q_0=0.0$ for
\citealt{sm96}, \citealt{ks98}, and \citealt{goto05}; $L_{60} \ge
10^{10.77}$L$_\odot$ with $H_0$= 100 km s$^{-1}$Mpc$^{-1}$ and
$q_0=0.5$ for \citealt{cl96a}; $L_{60} \ge 10^{12}$L$_\odot$ with
$H_0$= 50 km s$^{-1}$Mpc$^{-1}$ and $\Omega $=1 for
\citealt{law99}). In order to check the reliability of our
selection criteria of ULIRGs, we compiled 636 known ULIRGs taken 
mostly from wide area survey data
\citep{leech94,clowes95,mur96,cl96a,duc97,ks98,cl99,law99,rig99,stanford00,san03,goto05,cao06}.
We then applied the same procedure given in Section \ref{identi} to
select ULIRGs in this sample of known ULIRGs. Of these known
ULIRGs from the literature, we found 291 reliable sources in step
\ref{crit2} of Section \ref{identi}. A total of 345 known
ULIRGs were not found since they were neither observed in the
redshift surveys of this study nor included in FSC92. 
After the final identification step in Section \ref{identi}, we found 131
known ULIRGs in our final sample of ULIRGs, and failed to find a total
of 160 ULIRGs.
The reasons our having missed these 160 known ULIRGs are categorised as
follows:

\begin{enumerate}
\item {\bf Different redshifts} (F00415$-$0737, F01031$-$2255,
F01082$-$2452, F03014$-$2026, F03193$-$2224, F03448$-$2628,
F08509$-$1504, F10122+4943, F20087$-$0308, F20109$-$3003,
F20175$-$4756, F23515$-$2917): For the sources in this category,
the redshifts used in this study are different to (and almost always
much lower than) those in the literature. Therefore, the IR luminosities
calculated in this study are not large enough to be selected as
ULIRGs in this study. Some sources have high redshift qualities
(F00415$-$0737, F08509$-$1504, F10122+4943, F20109$-$3003), while
the redshift qualities for the others are low. The redshift of
F01031-2255 used in this study is larger than that in the
literature, but it is not included in the final sample of ULIRGs
due to its low redshift quality.

\item {\bf Different optical counterpart galaxies} (F03202$-$0001,
F08007+3928, F09346+3911, F12527$-$0306, F14390+6209, F14475+0155,
F14546+0338, F15182+3023, F21368+1006, F22011+0017,
F23051$-$0100): The optical counterpart galaxies of these {\it
IRAS} sources in this study are different from the galaxies in the
literature, as are their redshifts.

\item {\bf Different 60$\umu$m fluxes} (F00444$-$1803, F09320+6134,
F11087+5351, F14170+4545, F14351$-$1954, F14575+3256,
F18520$-$5048): The IR luminosities of these sources are not
greater than 10$^{12}$ L$_\odot$ since the 60 $\umu$m fluxes of these
sources used in this study (from FSC92) are lower than the fluxes
used in the literature (mostly from PSC). 

\item {\bf Marginal ULIRGs} (F12495$-$3414, F13156+0435, F14378$-$2604, F22509$-$0040): Since
the $L_{60}$ luminosities of all four sources are slightly less
than $10^{12}$L$_\odot$, they were classified as {\it further
marginal/possible ULIRGs} in \citet{law99}. Similarly, their IR
luminosities calculated in this study are in the range of
$10^{11.84}-10^{11.95}$ L$_\odot$, therefore they are not
classified as ULIRGs in this study.

\item {\bf Low qualities of redshifts} (F00406$-$3127, F10026$-$0022,
F14207$-$2002, F14485$-$2434, F22123$-$2025): These sources
satisfy the selection criteria up to step \ref{crit3} of Section
\ref{identi}. However, they fail to satisfy the criteria in step
\ref{crit4} of Section \ref{identi} due to their low redshift
qualities. Although the redshift qualities of F00406$-$3127,
F14485$-$2434 and F22123$-$2025 are low, these redshifts 
are same as those quoted in the literature. Therefore we include
these three sources in the final sample of ULIRGs. F14207$-$2002
is not a known ULIRG, but its redshift in this study is same as
that of \citet{allen91}. Therefore we also include F14207$-$2002
in the final sample of ULIRGs. However, F10026$-$0022 was
identified as a ULIRG in \citet{goto05} using the same SDSS data
with low quality of redshift as in this study. Therefore we reject
it from the final catalogue of ULIRGs.

\item {\bf Underluminous ULIRGs} (F00090$-$0054, F08112+3039, F08322+3609,
F09045+3943, F16134+2919, F21341$-$0033): These sources are the
FIRST-FSC sample of \citet{stanford00}. They did not use a strict
minimum infrared luminosity to identify ULIRGs. The far-infrared
(FIR) luminosities for these sources in \citet{stanford00} are in
the range of $10^{11.27}-10^{11.85}$ L$_\odot$ which is similar to
the minimum FIR luminosity of ULIRGs in \citet{sm96}. Similarly,
these sources are not selected as ULIRGs in this study due to
our low estimates for their 
IR luminosities ($10^{11.44}-10^{11.99}$ L$_\odot$).

\item {\bf Different SEDs} (F21219$-$1757): 
This source was classified as a ULIRG in
\citet{ks98} with log(L$_{ir}$/L$_\odot$)=12.06. However, the IR
luminosity calculated in this study is
log(L$_{ir}$/L$_\odot$)=11.86 which does not satisfy the ULIRGs
criteria, although the redshift and {\it IRAS} flux densities used
in this study are same as those in \citet{ks98}. This is because
we used M82 starburst SED to calculate the IR luminosity which may
not be appropriate for low luminosity QSOs such as this. If we use
AGN dominated SED\footnote{If we use AGN dominated SED of NGC 1068 to calculate the
IR luminosities for `{\it Warm}' ULIRGs ($f_{25}/f_{60}>$
0.2) in the final sample, the IR luminosities for those sources 
will be increased by 82 per cent.} of NGC 1068
to calculate the IR luminosity for this source,
we obtain the IR luminosity of log(L$_{ir}$/L$_\odot$)=12.14, satisfying
the ULIRGs criteria.

\item {\bf Using upper limit flux} (115 objects):
These sources were classified as ULIRGs in \citet{goto05} and
\citet{cao06}. Although most {\it IRAS} galaxies have upper limit
fluxes at 12$\umu$m and 25$\umu$m (to be discussed in the end of
this section), \citet{goto05} calculated the IR luminosities of 
{\it IRAS} sources by treating the flux upper limits as detections.
Similarly, three
sources for which 100$\umu$m flux densities are ``upper limits'' 
were found to be ULIRGs in \citet{cao06} by assuming the 
100$\umu$m upper limit is obtained. These calculations
can clearly result in overestimates for 
the IR luminosities 
of {\it IRAS} galaxies. These sources are not
selected in our study as ULIRGs due to their low IR luminosities of
$10^{10.60}-10^{11.99}$ L$_\odot$.

\end{enumerate}

In summary, all 160 known ULIRGs that are not identified in this
study have appropriate reasons not to be selected as ULIRGs except
for only one source (F21219$-$1757). Therefore we conclude that
our ULIRGs selection criteria are reliable and good enough to
identify nearly all known ULIRGs.

\begin{figure}
\begin{center}
\includegraphics [width=45mm] {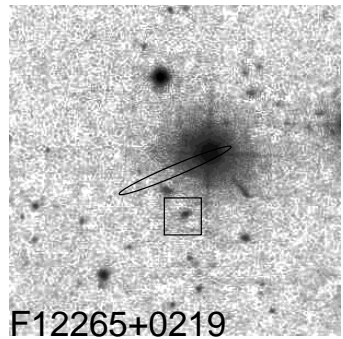}
\caption{The 2$'\times$2$'$ finding chart of F12265+0219 extracted
from the SDSS $gri$ composite image. The finding chart centred on
{\it IRAS} source position. North is up and East is to
the left. The ellipse represents 3$\sigma$ positional uncertainty,
and the square denotes the optical counterpart found in
\citet{goto05}, SDSS J122907.3+020246. The star like object within
the ellipse is 3C 273.} \label{fig-F12265+0219}
\end{center}
\end{figure}

\citet{goto05} performed a cross-correlation of FSC92 with the
spectroscopic catalogues of galaxies in the SDSS Third Data Release
(SDSSDR3), and identified 178 ULIRGs and 3 HLIRGs. Since the
SDSSDR4, which we used for our cross-correlation in this study,
contains galaxies of SDSSDR3, we can compare our identification
procedure of ULIRGs and HLIRGs with that of \citet{goto05}. Firstly,
we accept the correlation if a galaxy in SDSSDR4 lies within 3
$\sigma$ positional uncertainty ellipse of the {\it IRAS\/}
source, while \citet{goto05} treats a neighbouring optical galaxy
as a match if the galaxy in
SDSSDR3 lies within 20 arcsec of the {\it IRAS} source without
regard to the {\it IRAS} error ellipse. This can cause many problems. 
For example, SDSS J122907.3+020246 was identified as a HLIRG
in \citet{goto05}. In Figure \ref{fig-F12265+0219}, we present the
SDSS $gri$ composite image of F12265+0219, which was regarded as
the {\it IRAS} counterpart of SDSS J122907.3+020246 in
\citet{goto05}. 
It is seen that SDSS J122907.3+020246 lies outside the 
3$\sigma$ error ellipse
and a bright star-like object lies within the ellipse. The star-like 
object within the ellipse is the famous quasar 3C 273, which is
known to be associated with F12265+0219. Since 3C 273 was not
included in the spectroscopic targets of the SDSS and
\citet{goto05} used circular matching tolerance of 20 arcsec, the
mismatch between F12265+0219 and SDSS J122907.3+020246 was
included in his sample. Our cross-correlation based on the
positional error ellipse, and the NED Near Position Search in step
\ref{crit5} of Section \ref{identi} help to avoid this kind of
mismatching.
Secondly, there is a difference in the calculation of infrared
luminosity between \citet{goto05} and this study. The former used the same
function as used by \citet{ks98} for the calculation of IR
luminosity using {\it IRAS}
flux directly from FSC92. However, a significant fraction 
(81 per cent of ``high'' or ``moderate'' 60 $\umu$m sources for
$f_{12}$ fluxes\footnote{The quantities $f_{12}$, $f_{25}$,
$f_{60}$, and $f_{100}$ represent the {\it IRAS} flux densities in
Jy at 12$\umu$m, 25$\umu$m,
60$\umu$m, and 100$\umu$m, respectively.} and 74 per cent for $f_{25}$ fluxes) 
of the {\it IRAS} sources only have upper limits for $f_{12}$ and
$f_{25}$ fluxes. Therefore, assuming upper limits are obtained for
the measurements at $f_{12}$ and
$f_{25}$ in calculating the IR luminosity can lead to overestimates of
the IR luminosity, although the
uncertainty in the IR luminosity is only a few percent
\citep{ks98}.

\subsection{Reliability and Completeness of the final sample}\label{rel_sample}
\begin{figure}
\begin{center}
\includegraphics [width=84mm] {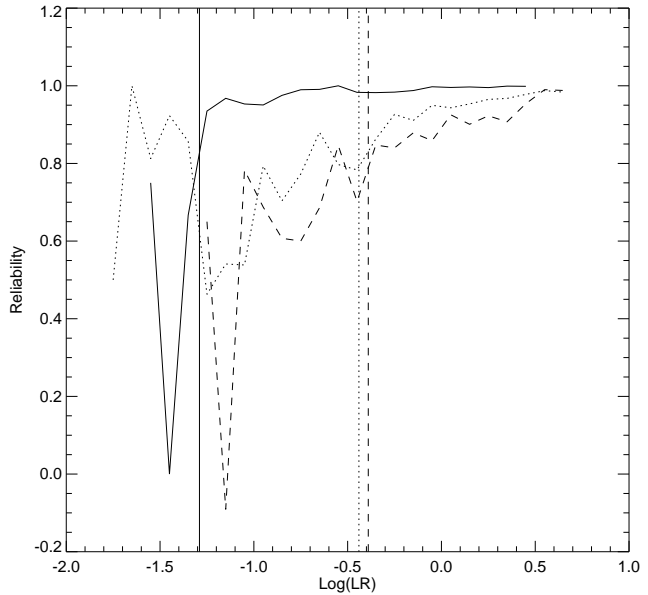}
\caption{Reliability of our cross-correlation for SDSSDR4 (dotted
line), 2dFGRSF (dashed line) and 6dFGSDR2 (solid line) as a
function of LR. The vertical lines represent the critical LR
values of reliable identification for each redshift survey data -
above the critical LR values, the reliabilities are $\geq$ 80 \%.}
\label{fig-rel}
\end{center}
\end{figure}
A final sample of ULIRGs selected by cross-correlation based on
the position alone may contain spurious sources due to the chance
presence of a galaxy within the {\it IRAS} 3 $\sigma$ ellipse. In
order to estimate the probability of the ``true" association
between {\it IRAS} sources and the optical counterparts, we
compute likelihood ratio (\citealt{sutherland92}) for each
association. The likelihood ratio defined by the ratio of the
probability of a true association to that of a chance association
is\footnote{We assume that a positional error of {\it IRAS} source
is a Gaussian.}
\begin{equation}
{\rm LR} = \frac{Q(\leq m) \exp(-r^{2}/2)}{2\pi\sigma_{\rm
1}\sigma_{\rm 2}n(\leq m)}, \label{eq-LR}
\end{equation}
where $n(\leq m)$ is the local surface density of objects brighter
than the candidate. The ``normalised distance" $r$ is given by
\begin{equation}
r^{2} = \left( \frac{d_1}{\sigma_{1}} \right)^{2} +\left(
\frac{d_2}{\sigma_{2}} \right)^{2}, \label{eq-normdist}
\end{equation}
where the $d_1$, $d_2$ are the positional differences along the
two axes of an error ellipse for an {\it IRAS} source, and
$\sigma_1$, $\sigma_2$ are the lengths of these axes. Since the
positional uncertainties of galaxies in SDSS, 2dFGRS, and 6dFGS
are negligible compared with those of {\it IRAS} sources, we define
$\sigma$ as 
the length of the error axes of the {\it IRAS} sources. 
$Q(\leq m)$ is a multiplicative factor which is the a priori
probability that a ``true'' optical counterpart brighter than the
magnitude limit exists in the association, and for simplicity we
set $Q=1$ in this study.

\begin{figure}
\begin{center}
\includegraphics [width=84mm] {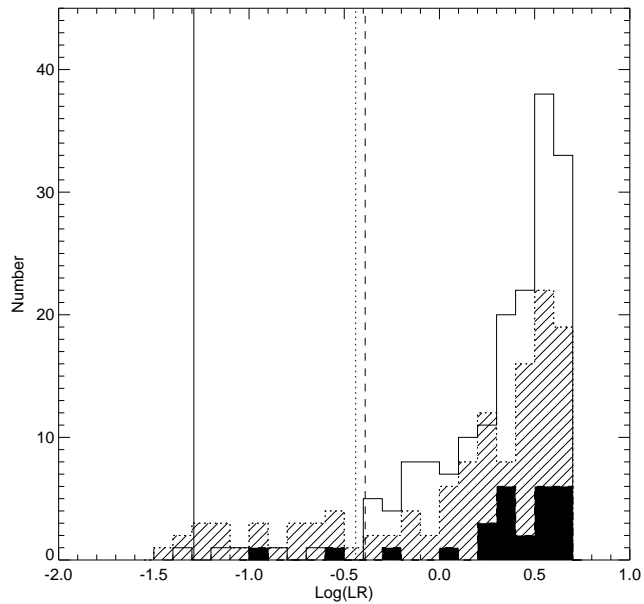}
\caption{Distribution of LR values for the final sample of 324
ULIRGs found in SDSSDR4 (shaded region), 2dFGRSDRF (filled
region), and 6dFGSDR2 (solid line). The vertical lines represent
the critical LR values of reliable identification (the reliability of $\sim$ 80 \%) as shown in
Figure \ref{fig-rel}.} \label{fig-lrdist}
\end{center}
\end{figure}

We compute $n(\leq m)$ using photometric sample of galaxies within
3 $\sigma$ error ellipse,
\begin{equation}
n(\leq m) = \frac{N(\leq m)}{9\pi \sigma_{\rm 1}\sigma_{\rm 2}}
\label{eq-density}
\end{equation}
where $N(\leq m)$ represents the number of galaxies of which
magnitudes\footnote{We use SDSS Petrosian r-band magnitude for the
galaxies in SDSSDR4, and ${\rm b_J}$ band magnitude for the
galaxies in 2dFGRSDRF and 6dFGSDR2.} are less than or equal to
that of a candidate. We then obtain the likelihood ratio for our
sample
\begin{equation}
{\rm LR} = \frac{9\exp(-r^{2}/2)}{2N(\leq m)}. \label{eq-LRfin}
\end{equation}
We compute LR values for all 19,380
sources in the list of all matched sources of Section \ref{sample}
using the photometric sample of galaxies in each redshift survey. In
order to calculate the reliability of association using the LR values,
we perform random associations by offsetting the {\it IRAS} source
positions by $\approx30'$, and recompute LR values for each random
association following \citet{lonsdale98} and \citet{masci01}.
Using the distribution of LR values for true associations and
random associations, the reliability of each association with a
given LR is defined by
\begin{equation}
R({\rm LR}) = 1 - \frac{N_{\rm random}({\rm LR})}{N_{\rm
true}({\rm LR})}, \label{eq-rel}
\end{equation}
where $N_{\rm true}({\rm LR})$ and $N_{\rm random}({\rm LR})$ are
the number of true and random associations with a given LR. In
Figure \ref{fig-rel}, we present the reliability of our
cross-correlation for each redshift survey as a function of LR.
Above log (LR) $\sim-$0.44 (SDSSDR4), $-$0.39 (2dFGRSDRF), $-$1.29
(6dfGSDR2), the reliabilities are $\geq$ 80\%. We show the LR
distribution of our ULIRG sample in Figure \ref{fig-lrdist} along
with critical LR values of reliable identification for each
redshift survey. If we apply the critical LR values of 80\%
reliability, then 24 out of 126 (SDSSDR4), 2 out of 27
(2dFGRSDRF), and 1 out of 171 (6dFGSDR2) ULIRGs can be rejected
from a final sample of ULIRGs. 

The reliability of our LR-based identifications is 
confirmed by the NED Near Position Search in Section \ref{identi}.
Of the 11 objects that were rejected from our final sample of ULIRGs
because the optical association differs from that in NED, eight 
have lower LR values than the critical values. Although three
objects have larger LR values than the critical values, there are
more appropriate galaxies in NED (the larger value of LR) that would
result in the galaxies not being ULIRGs.

The spectroscopic target selection in a redshift survey introduces
incompleteness into our final sample of ULIRGs. 
Since the target selection
functions differ in our three redshift surveys (a short
discussion is presented in Section \ref{comp_sub}) and are beyond
the scope of this paper, we only estimate the incompleteness
originated from this study. Therefore, we measure the completeness
relative to all FSC92 ULIRGs that are identified in these surveys, 
and {\it not} for example relative to a volume-limited ULIRG survey
or to FSC92 as a whole.  

There are three factors which make our
sample incomplete. Firstly, since we used 3 $\sigma$ error ellipse
for our cross-correlation, some ULIRGs outside 3 $\sigma$ error
ellipse might be missed. If we assume an error ellipse to be a
Gaussian distribution, the incompleteness originated from this
term would be about 1 per cent. Secondly, some ULIRGs for which
redshift qualities are unreliable (step \ref{crit4} in Section
\ref{identi}) are rejected, although their redshifts are not
wrong. Thirdly, when more than two galaxies including a known
optical counterpart galaxy are within the error ellipse of one {\it
IRAS} source, the wrong galaxy may be selected as the optical
counterpart of the {\it IRAS} source. In order to quantify the
incompleteness introduced by those factors, we compute the ratio
of ULIRGs that are surveyed in SDSSDR4, 2dFGRSDRF and 6dFGSDR2 to
those that are missed from our final sample of ULIRGs using
previously known ULIRGs. Of the 636 known ULIRGs in Section
\ref{rel}, we consider only the 140 known ULIRGs that are expected
to be found in this study. This sample consists of ULIRGs that are
located within the redshift survey region and are found in FSC92.
Additionally, their IR luminosities calculated using our method
are greater than 10$^{12}$ L$_\odot$. In the result of
identification procedure in Section \ref{sample} and \ref{identi},
133 ULIRGs are found in our final sample of ULIRGs, and 7 ULIRGs
are missed. 

Of the 7 missed ULIRGs, three {\it IRAS} sources
(F03202-0001, F14546+0338 and F22011+0017) have more than two
galaxies including a known optical counterpart within their error
ellipses. Our selection procedure is to choose the most likely 
counterpart (i.e. the galaxy with the larger value of LR), 
but this selects different galaxies to the known
optical counterparts for these three {\it IRAS} sources. 
Therefore, the incompleteness introduced by our selection
using LR values among several counterpart candidates is about 2.1 per cent (3 out of 140),
although all the optical counterparts that we select are brighter or closer
to the {\it IRAS} position than the known optical counterparts. 
In addition, one ULIRG
(F09111-1007) is missed, because the optical counterpart in
6dFGSDR2 is outside 3 $\sigma$ error ellipse, although the
association is already confirmed in the literature. Three ULIRGs
(F00406$-$3127, F14485$-$2434, and F22123$-$2025) are missed
because of their low redshift qualities as shown in Section \ref{rel},
although the redshifts quoted in the redshift surveys are same as
those in the literature. Therefore, the incompleteness
introduced by our cross-correlation algorithm
using 3 $\sigma$ error ellipse
is about 0.7 per cent (1 out of 140), and the incompleteness 
introduced by our selection
using the redshift quality parameter in the step \ref{crit4} of
Section \ref{identi} is about 2.1 per cent (3 out of 140).
In total, the incompleteness of the final sample of ULIRGs
originated from this study is about 5 per cent (7 out of 140).

\section{Results}

We have found 324 ULIRGs (126 in SDSSDR4, 27 in 2dFGRSDRF, and 171
in 6dFGSDR2), 190 of which are newly discovered.
We have therefore increased the number of known ULIRGs by about 30 per
cent. 

\begin{figure}
\includegraphics [width=84mm] {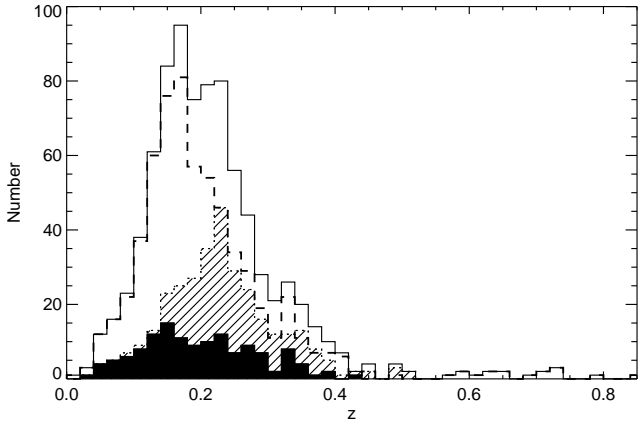}
\caption{Redshift distribution of the 324 ULIRGs identified in this
study (shaded region) compared to the 636 known ULIRGs (dashed
lines), 134 known ULIRGs in the final sample in this study (filled
region) and known ULIRGs plus all identified ULIRGs in this study
(solid lines).}\label{fig-zdist}
\end{figure}

In Figure \ref{fig-zdist}, we present the redshift distribution
for the final sample of 324 ULIRGs identified in this study, and
compare it to that of the 
636 previously known ULIRGs. The redshifts in our final sample run
from z=0.037 to z=0.517 and the median value is $\bar{z}$=0.223,
which is larger than the median redshift $\bar{z}$=0.184 of the 
636 previously
known ULIRGs. 

\begin{figure}
\includegraphics [width=84mm] {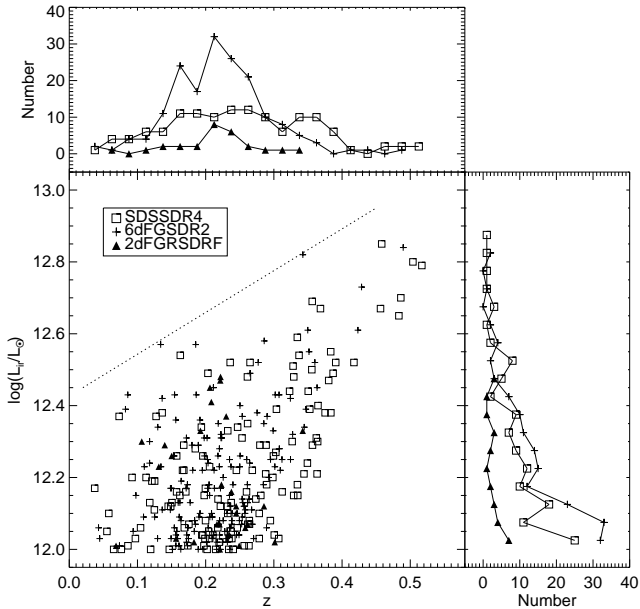}
 \caption{The IR luminosity of identified
ULIRGs in this study versus redshift. The ULIRGs from the SDSSDR4,
2dFGRSDRF and 6dFGSDR2 are represented by squares, triangles, and
pluses, respectively. In the upper and right panel, the redshift
and luminosity distribution of ULIRGs found in each redshift
survey are shown, respectively.} \label{fig-zlum}
\end{figure}

In Figure \ref{fig-zlum}, we plot the IR luminosities of
the ULIRGs identified in this study against their redshifts. 
In the upper and
right panel, we present the redshift and luminosity distribution
of ULIRGs found in each redshift survey, respectively.
The ULIRGs identified in SDSSDR4 are found from low z to high z
and from low luminosity to high luminosity ($\sim 10^{12.85}
L_\odot$). The ULIRGs in 6dFGSDR2 show similar distribution with
those in SDSSDR4, although median redshift of galaxies in 6dFGSDR2
is smaller than that of galaxies in SDSSDR4. In contrast, in spite
of relatively larger median redshift of galaxies in 2dFGRSDRF
compared to that of 6dFGSDR2, no ULIRGs in 2dFGRSDRF are found
beyond z$\sim$0.35. This might be due to the different selection
criteria for spectroscopic targets (see each data release paper
for more detail).
Figure \ref{fig-zlum} shows that the upper envelope (shown by a
dotted line) increases from log $(L_{ir}/L_\odot) = 12.45$ to
$12.95$ as redshift increases. The lower envelope is due to the
survey detection limit.  

\begin{figure}
\includegraphics [width=84mm] {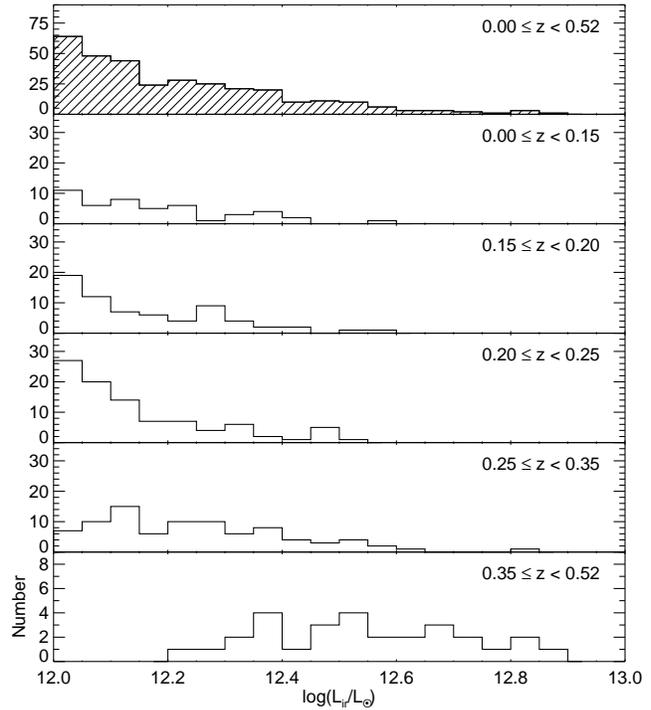}
\caption{The luminosity distribution for the final sample of 324
ULIRGs is plotted in the top panel. Below this panel, the
luminosity distributions of ULIRGs in the redshift ranges
$0.00\leq z <0.15$, $0.15\leq z <0.2$, $0.2\leq z <0.25$,
$0.25\leq z <0.35$, and $0.35\leq z$ are shown.} \label{fig-ldist}
\end{figure}

The luminosity distribution of the final sample is plotted in the
Figure \ref{fig-ldist}. For the final sample of ULIRGs (top
panel), there are fewer ULIRGs at high luminosities than at 
low luminosities. The median luminosity value is
10$^{12.17} L_\odot$. Two interesting features are seen in the
luminosity variation with redshift in Figure \ref{fig-ldist}.
Firstly, the number ratio of low luminous ULIRGs (L$_{ir}< 10^{12.2}
L_\odot$) to all ULIRGs in redshift bin increases with redshift
up to z=0.25 (due to the detection limit seen in
Figure \ref{fig-zlum}, there are few low luminous ULIRGs beyond
z=0.25). Secondly, no ULIRGs that have IR luminosities of log
$(L_{ir}/L_\odot) > 12.60$ are found in the lower redshift
(z$<$0.25). This is almost certainly due to the evolving 
luminosity function. A calculation of the evolving luminosity
function is beyond the scope of this paper.

\section{Discussion}
\subsection{Comparison of ULIRG subsamples}\label{comp_sub}
\begin{figure}
\includegraphics [width=84mm] {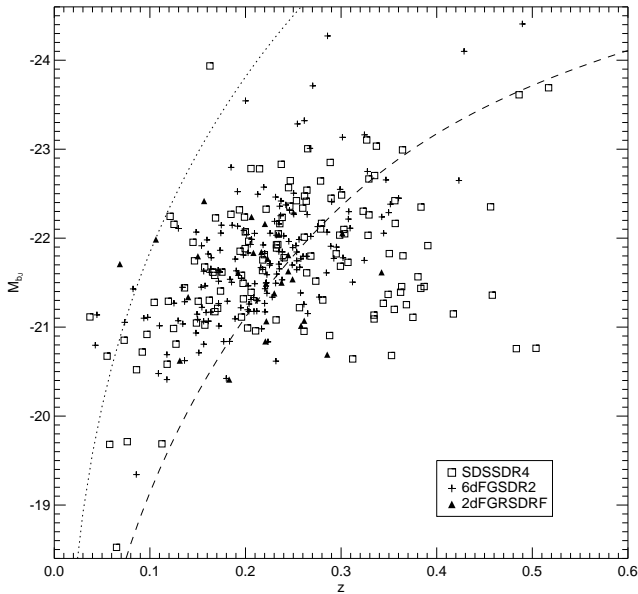}
\caption{Absolute ${\rm b_J}$ magnitude of ULIRG subsamples
against redshift. The ULIRGs from the SDSSDR4, 2dFGRSDRF and
6dFGSDR2 are represented by squares, triangles, and plusses,
respectively. The dotted and dashed line indicate the ${\rm b_J}$
magnitude limit of 6dFGSDR2 and 2dFGRSDRF, respectively.}
\label{fig-absmag}
\end{figure}
Since we use three different redshift survey data to search for
ULIRGs, different target selection functions in each redshift
survey make our final sample of ULIRGs inhomogeneous. The majority
of spectroscopic targets in each redshift survey are selected
from  magnitude limited 
samples\footnote{The primary sample ($\sim 70\%$) of 6dFGS
are galaxies with $K_{\rm tot}<$12.75 and one additional
sample is optically selected galaxies with ${\rm
b_J}<16.75$~($\sim 5\%$). Most spectroscopic targets of 2dFGRS are
galaxies with ${\rm b_J}<19.45$, and the main galaxy sample of
SDSS are galaxies with Petrosian r-band magnitudes brighter than
r=17.77.}, but some additional spectroscopic targets are included
depending on the redshift survey. In order to see the effects of
the magnitude limit of each survey on our ULIRG subsamples, we plot
the absolute ${\rm b_J}$ magnitude against redshift 
of ULIRG subsamples in Figure \ref{fig-absmag}. A transformation
from SDSS photometry to ${\rm b_J}$ magnitude, and k-correction is
done following \citet{norberg02}. Since the
majority of the spectroscopic targets in 6dFGS are $K$-band
selected, most ULIRGs found in 6dFGSDR2 are fainter than the
${\rm b_J}$ magnitude limit (${\rm b_J}=16.75$) shown by the dotted
line. The majority of ULIRGs found in 2dFGRSDRF are above the
magnitude limit (${\rm b_J}=19.45$). The ${\rm b_J}$ magnitude
limit for the
spectroscopic sample of SDSSDR4 is not clearly defined since the
main spectroscopic sample of galaxies are Petrosian r-band
selected sample. However, ${\rm b_J}$ magnitude limit for the main
galaxy sample of SDSSDR4 is similar to that of 2dFGRSDRF,
and most galaxies fainter than the ${\rm b_J}$ magnitude limit are
``Luminous Red Galaxies'' (see \citealt{am05} for more detail). It
appears that a significant fraction (37\%) of ULIRGs found in
SDSSDR4 are fainter than the magnitude limit of 2dFGRS, implying 
that they are Luminous Red Galaxies.

In order to compare the infrared properties of our ULIRG
subsamples associated with each redshift survey, we plot the
infrared colour of log($f_{25}/f_{60}$) against
log($f_{60}/f_{100}$) (a detailed discussion on the infrared colours
of ULIRGs will be given in Section \ref{ircolors}). Although the
majority of ULIRGs in the range of log($f_{60}/f_{100}$)$<-$0.4
or in the range of log($f_{25}/f_{60}$)$>-$0.7 are {\it IRAS} sources with flux
quality of 1 at 100$\umu$m or at 25$\umu$m (see Figure
\ref{fig-colour2} and \ref{fig-colour3}), there are no striking
differences in infrared colours among the ULIRG subsamples
associated with each redshift survey.

\begin{figure}
\includegraphics [width=84mm] {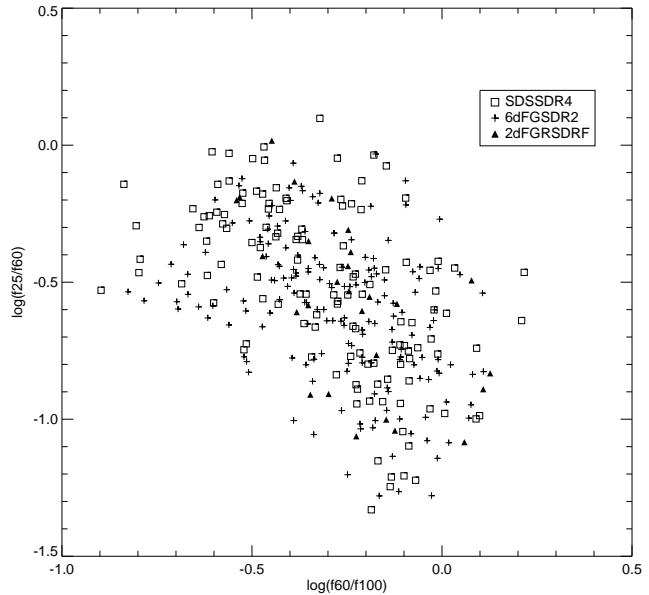}
\caption{Infrared colour-colour diagram. Log($f_{25}/f_{60}$) 
is plotted against
log($f_{60}/f_{100}$) for the ULIRG subsamples. The ULIRGs from the
SDSSDR4, 2dFGRSDRF and 6dFGSDR2 are represented by squares,
triangles, and plusses, respectively.} \label{fig-ircols}
\end{figure}
In conclusion, the sample of ULIRGs found in the three redshift
surveys are not a simple magnitude limited sample due to the 
diverse target selection functions of each redshift survey.

\subsection{Infrared Colours}\label{ircolors}
Infrared colours such as log($f_{12}/f_{60}$),
log($f_{25}/f_{60}$) and log($f_{60}/f_{100}$) have long
been used for
classifying infrared sources. Infrared-bright stars 
have SEDs peaking typically 
around 12 $\umu$m, and can be distinguished from
galaxies using the colour of log($f_{12}/f_{60}$) (see, e.g.,
\citealt{cohen87}). In addition, 
log($f_{25}/f_{60}$) has been used to classify ULIRGs into `{\it Warm}'
and `{\it Cool}' systems. It has been suggested that ULIRG may evolve
from `{\it Cool}' system to `{\it Warm}' system
\citep{san88b,vks02}. \citet{sn91} showed that for luminous
infrared galaxies including ULIRGs in the {\it IRAS} Bright Galaxy
Survey, the mean log($f_{60}/f_{100}$) colour increases with
increasing IR luminosity.

\begin{figure}
\includegraphics [width=84mm] {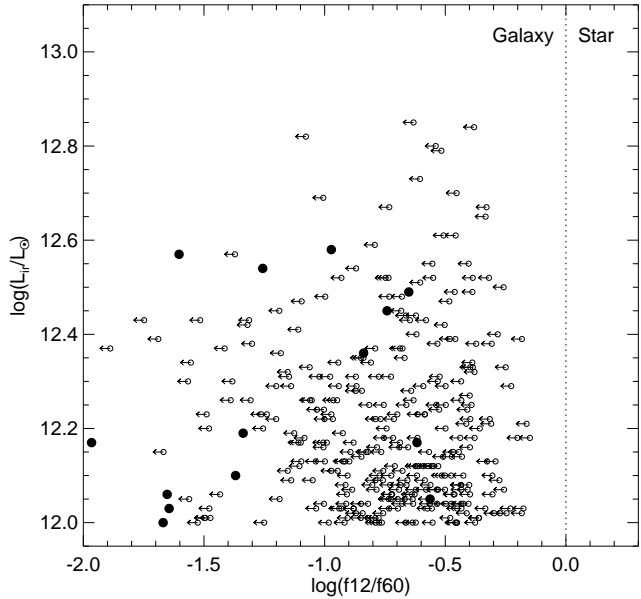}
\caption{The infrared luminosity against the infrared colour
log($f_{12}/f_{60}$) for the final sample of 324 ULIRGs in this
study. The ULIRGs with flux upper limits at 12 $\umu$m are
represented by open circles with arrows indicating the sense of
the limit, and the ULIRGs with high or moderate flux qualities are
represented by filled circles. The dotted line represents the
boundary between galaxies and stars.} \label{fig-colour1}
\end{figure}

\begin{figure}
\includegraphics [width=84mm] {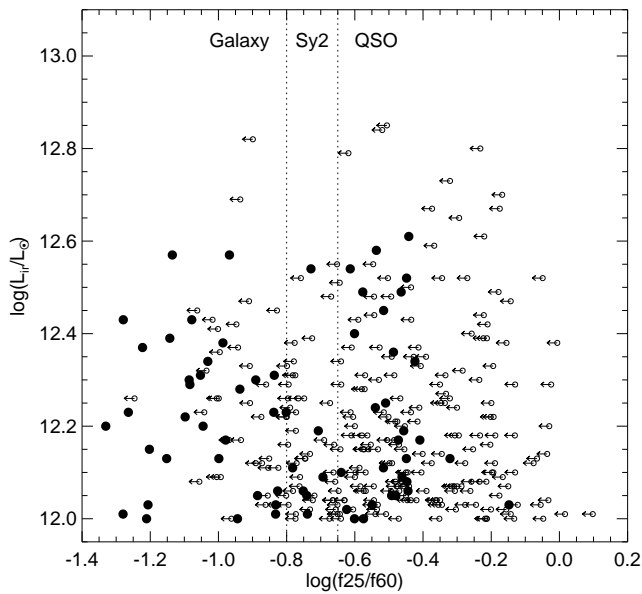}
\caption{As Figure \ref{fig-colour1}, except using the 
infrared colour log($f_{25}/f_{60}$). The open circles and arrows
represent the ULIRGs which have upper limits at 25 $\umu$m,
and the filled circles denote the ULIRGs which have high or
moderate flux qualities at 25 $\umu$m. The dotted lines represent
the boundaries among normal galaxies, Seyfert 2 galaxies, and
QSOs.} \label{fig-colour2}
\end{figure}

\begin{figure}
\includegraphics [width=84mm] {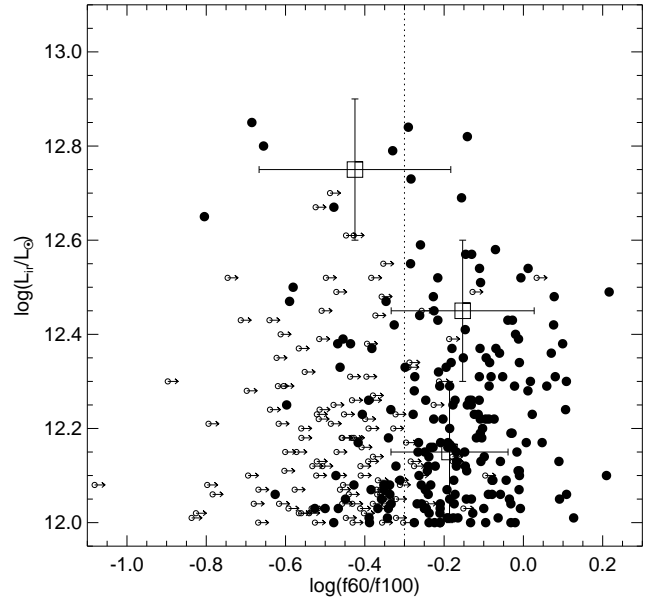}
\caption{As Figure \ref{fig-colour1}, except using the 
infrared colour log($f_{60}/f_{100}$). The ULIRGs with flux upper
limits at 100 $\umu$m are represented by open circles with arrows
indicating the sense of the limit, and the ULIRGs with high or
moderate flux qualities are represented as filled circles. The
open squares are at the mean colour and the central luminosity of
ULIRGs in the three luminosity bins ($12.0\leq log(L_{ir}/L_\odot)
<12.3$, $12.3\leq log(L_{ir}/L_\odot) <12.6$, and $12.6\leq
log(L_{ir}/L_\odot) <12.9$). The vertical errorbars define the
limiting luminosities of the bins, and the horizontal errorbars
represent standard deviation of log($f_{60}/f_{100}$) colours in
each bin. The dotted line represents the selection criteria
(log($f_{60}/f_{100}$)$>-$0.3) of 1 Jy sample of ULIRGs
\citep{ks98}.} \label{fig-colour3}
\end{figure}

We present the IR luminosity against infrared colour of
log($f_{12}/f_{60}$) in Figure \ref{fig-colour1}, against
log($f_{25}/f_{60}$) in Figure \ref{fig-colour2}, and against
log($f_{60}/f_{100}$) in Figure \ref{fig-colour3} for the final
sample of 324 ULIRGs. The infrared-bright stars are known to have
log($f_{12}/f_{60}$)$>0$ (see, e.g., \citealt{ks98}). 
All ULIRGs in the final sample
have infrared colours of log($f_{12}/f_{60}$)$<0$.

In Figure \ref{fig-colour2}, we plotted the boundary lines for the
classification of normal galaxies, Seyfert 2 galaxies, and QSOs
suggested by \citet{nh92}. If we restrict our attention to the ULIRGs 
with flux qualities of 2 or 3 (filled circles), the number ratio of these
objects is N(Galaxy) : N(Sy2) : N(QSO) $= 32 : 7 : 29$. If we use
the classification scheme of `{\it Warm}' ULIRGs ($f_{25}/f_{60}>$
0.2, or log($f_{25}/f_{60}$)$>-$0.7) and `{\it Cool}' ULIRGs
($f_{25}/f_{60}<$ 0.2, or log($f_{25}/f_{60}$)$<-$0.7) suggested
by \citet{san88b}, the number ratio for the sample becomes
N($Cool$) : N($Warm$) $= 38 : 30$. This ratio is much smaller than
that of 1 Jy sample (N($Cool$) : N($Warm$) $= 90 : 25$,
\citealt{ks98}). This difference might be related to the different
selection criteria for ULIRGs, but is likely to be strongly affected
by the exclusion of ULIRGs which
have upper limit flux at 25$\umu$m.

In Figure \ref{fig-colour3}, the log($f_{60}/f_{100}$) colours of
ULIRGs with flux quality of 2 or 3, range from $-$0.80 to +0.22
and have a mean of $-$0.19. This mean value is low 
compared to those of other ULIRG samples (see, e.g.,
\citealt{ks98} and references therein). In addition, 21 per cent
(42 out of 203) of ULIRGs with flux quality of 2 or 3, have
colours of log($f_{60}/f_{100}$) less than $-$0.3, while no ULIRGs
in 1 Jy sample have colours of log($f_{60}/f_{100}$) less than
$-$0.3 due to their selection criteria
(log($f_{60}/f_{100}$)$>-$0.3) of ULIRGs \citep{ks98}. We plotted
the mean colours in each luminosity bin by open squares to
investigate any dependence of the colour on the IR luminosity. It
appears that there are no significant IR luminosity dependence 
of log($f_{60}/f_{100}$) in our
sample (up to $10^{12.6} L_\odot$), though our colours are still
warmer than those typical in less IR-luminous galaxies. 

We compare the IR luminosity and the redshift distribution of ULIRGs with flux qualities
of 1 (open circles in Figure \ref{fig-colour1}, \ref{fig-colour2}, and \ref{fig-colour3}) 
to those with flux qualities of 2 or 3 (filled circles). 
The mean IR luminosity for ULIRGs with flux qualities
of 1 at 12 $\umu$m is not different from that for ULIRGs with flux qualities of 2 or 3 as seen in Figure \ref{fig-colour1}.
Similar results are found for ULIRGs at 25 $\umu$m and at 100 $\umu$m. The median redshift ($\bar{z}$=0.230)
for ULIRGs with flux qualities of 1 at 12 $\umu$m is larger than that ($\bar{z}$=0.163) for ULIRGs with flux qualities of 2 or 3 at 12 $\umu$m.
Similarly, the median redshifts for ULIRGs with flux qualities of 1 at 25 $\umu$m ($\bar{z}$=0.236) and at 100 $\umu$m ($\bar{z}$=0.262) 
are larger than those for ULIRGs with flux qualities of 2 or 3 at 25 $\umu$m ($\bar{z}$=0.148) and at 100 $\umu$m ($\bar{z}$=0.196).

\begin{table*}
\centering
\caption{The Final Sample of HLIRGs\label{tab-HLIRGs}}
\begin{tabular}{crrrrrrcccr}
\hline\hline FSC & RA & Dec & 12$\umu$m & 25$\umu$m & 60$\umu$m &
100$\umu$m & Flux &
final & log $L_{ir}$ &log(LR)  \\
NAME & (J2000) & (J2000) & (Jy) & (Jy) & (Jy) & (Jy)  & Qual&
z & ($L_\odot$) \\
\hline F01044$-$4050 & 01 06 44.9 & $-$40 34 21 & 0.140(0.042) &
0.155(0.039) & 0.405(0.049) & 0.655(0.203) & 1231 &
0.584 & 13.19 & 0.648\\
F09105+4108 & 09 13 44.0 & 40 56 34 & 0.129(0.031) & 0.333(0.033)
& 0.525(0.042) & 0.437(0.109) & 2331 &
0.442 & 12.97 & 0.191\\
\hline
\end{tabular}
\end{table*}

\begin{figure*}
\begin{center}
\includegraphics [width=148mm] {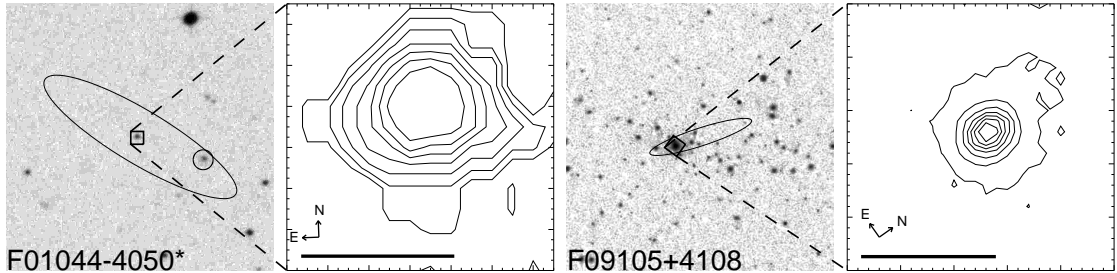}
\end{center}
 \caption{Similar to the Figure \ref{fig-ULIRGs}, except for two HLIRGs,
 F01044$-$4050 (left two panels) and F09105+4108 (right two
panels), identified in this study. The small circle in the first
panel indicate the position of 6dF J0106412$-$403437.}
\label{fig-HLIRGs}
\end{figure*}

\subsection{Hyperluminous Infrared Galaxies}\label{HLIRGs}

Of the 14 HLIRG candidates satisfying selection criteria of HLIRGs
up to step \ref{crit3} in Section \ref{identi}, all were excluded
in futher steps. However, we identified two HLIRGs, F01044$-$4050 and
F09105+4108, in the course of NED Near Position Search for ULIRGs
candidates at step \ref{crit5} in Section \ref{identi}. We list the 2
HLIRGs identified in this study in Table \ref{tab-HLIRGs}. Column
(1) lists the {\it IRAS\/} object name in the FSC92. Columns (2)
and (3) list the J2000.0 source position. Column (4), (5), (6) and
(7) give the {\it IRAS\/} flux densities (and their errors) at
12$\umu$m, 25$\umu$m, 60$\umu$m and 100$\umu$m, respectively. The
{\it IRAS\/} flux density quality at each band is given in Column
(8). Column (9) lists the finally accepted redshift and Column
(10) gives the IR luminosity calculated in this study. Log (LR) is
given in the final Column. In Figure \ref{fig-HLIRGs}, we present
3$'\times$3$'$ grayscale images and contour plots extracted from
the SuperCOSMOS scans of r$_F$ survey plates (left panel), as 
well as those from the SDSS $gri$ composite image and the SDSS r-band
image (right panel) for the final sample of two HLIRGs. The
identification procedure and properties of the HLIRGs are given in
Appendix B of the electronic issue.

\section{Summary}

We present a new sample of ULIRGs and HLIRGs found by
cross-correlating the {\it IRAS\/} sources in FSC92 with the
spectroscopic samples of galaxies in the SDSSDR4, 2dFGRSDRF and
6dFGSDR2. Our primary results are summarised below:

\begin{enumerate}
\item We have identified 324 ULIRGs including 190 newly discovered ULIRGs
in the regions of the sky covered by the SDSSDR4, 2dFGRSDRF and
6dFGSDR2. We increase the number of catalogued ULIRGs by about 30
per cent.

\item The reliability of the cross-correlation is estimated using 
the likelihood
ratio method. We compute the likelihood ratio of each
association for our sample of ULIRGs. The completeness of the final
sample has been estimated using previously known ULIRGs in the
redshift surveyed region. The incompleteness introduced by our
identification procedure due to the cross-correlation using 3
$\sigma$ error ellipse, and the selection using LR value and
using the redshift quality parameter is estimated to be about 5 per cent.

\item The redshifts in our final sample run from z=0.037 to z=0.517 and
the median value is $\bar{z}$=0.223, which is larger than that
($\bar{z}$=0.184) in previous ULIRG samples.

\item Two HLIRGs, F01044$-$4050 and F09105+4108, are found in the course of
NED Near Position Search of ULIRGs candidates.

\end{enumerate}

\section*{Acknowledgments}
We would like to thank Dr. Woong-Seob Jeong and Jung-Hoon Kim for
helpful discussions, and to anonymous referee for useful comments. 
H.S.H. and M.G.L. were supported in part by
ABRL (R14-2002-058-01000-0) and the BK21 program of the Korean
Government. K.H.L. acknowledges the support of the Royal Society
RM/Korea/XF12003/16356. 
We would also like to thank all the
people involved in creating the SDSS, 2dFGRS, 6dFGS,
SuperCosmos surveys, and NED.
Funding for the SDSS and SDSS-II has been provided by the Alfred P. Sloan Foundation, 
the Participating Institutions, the National Science Foundation, 
the U.S. Department of Energy, the National Aeronautics and Space Administration, 
the Japanese Monbukagakusho, the Max Planck Society, and the Higher Education Funding Council for England.
The SDSS Web Site is http://www.sdss.org/.
The SDSS is managed by the Astrophysical Research Consortium for the Participating Institutions. 
The Participating Institutions are the American Museum of Natural History, 
Astrophysical Institute Potsdam, University of Basel, University of Cambridge, 
Case Western Reserve University, University of Chicago, Drexel University, Fermilab, 
the Institute for Advanced Study, the Japan Participation Group, Johns Hopkins University, 
the Joint Institute for Nuclear Astrophysics, the Kavli Institute for Particle Astrophysics and Cosmology, 
the Korean Scientist Group, the Chinese Academy of Sciences (LAMOST), Los Alamos National Laboratory, 
the Max-Planck-Institute for Astronomy (MPIA), the Max-Planck-Institute for Astrophysics (MPA), 
New Mexico State University, Ohio State University, University of Pittsburgh, University of Portsmouth, 
Princeton University, the United States Naval Observatory, and the University of Washington.
This research has made use of the NASA/IPAC Extragalactic Database (NED) which is 
operated by the Jet Propulsion Laboratory, California Institute of Technology, 
under contract with the National Aeronautics and Space Administration.

\label{lastpage}

\end{document}